\documentclass[a4paper]{article}
\usepackage{upgreek}
\usepackage{graphicx}
\usepackage{subfig}
\usepackage{comment}
\usepackage{cancel}
\usepackage{paralist}
\usepackage{amsmath}
\usepackage{jinstpub} 

\author[a]{G. Bencivenni,}
\author[b]{G. Cibinetto,}
\author[c]{R. de Oliveira,}
\author[b]{R. Farinelli,}
\author[a]{G. Felici,} 
\author[a]{M. Gatta,}
\author[a,d]{M. Giovannetti,}
\author[b]{L. Lavezzi,}
\author[a,1]{G. Morello,\note{Corresponding author.}}
\author[e]{A. Ochi,}
\author[a]{M. Poli Lener,}
\author[a,f]{E. Tskhadadze} 

\affiliation[a]{Laboratori Nazionali di Frascati dell`INFN,\\Frascati, Italy}
\affiliation[b]{INFN Ferrara, \\Ferrara, Italy}
\affiliation[c]{CERN, \\Meyrin, Switzerland}
\affiliation[d]{University of Rome ``Tor Vergata'', \\Rome, Italy}
\affiliation[e]{Kobe University, \\Kobe, Japan}
\affiliation[f]{Technological Institute of Georgia, \\Tbilisi, Georgia}

\title{On the space resolution of the $\upmu$-RWELL}
\abstract{In MPGD detectors evaluation of the space resolution with the charge centroid (CC) method provides large uncertainty when the impinging particle is not perpendicular to the readout plane. An improvement of the position reconstruction, and thus of the space resolution, is represented by the $\upmu$TPC algorithm. In this work we report the application of this algorithm to the $\upmu$-Resistive WELL detector. Moreover a combination of the CC method with the $\upmu$TPC algorithm is proposed, showing an almost uniform resolution over a wide angular range.}

\keywords{Gaseous detectors; Micro-Pattern Gaseous Detectors; Gas Electron Multiplier; Resistive detectors; micro-Resistive WELL; Diamond-Like-Carbon; $\upmu$TPC mode.}
\begin{document}
\maketitle	

\section{Introduction}
Space resolution in MPGD can be affected by several factors: primary statistics, electrons diffusion in gas, readout geometry, Front-End Electronics (digital or analog FEE) and impinging angle $\uptheta$ of the crossing particle with respect to the normal to the readout electrode (fig.~\ref{sketch}). Indeed the larger is the angle the worse is the resolution $\upsigma_{x}$ usually evaluated with the charge centroid method. For an experiment this means a not uniform resolution in the solid angle covered by the apparatus and results that can be consequently characterized by large systematic errors. The first four factors are usually optimized with a dedicated R\&D on detector geometry, gas mixture and FEE while for the last factor a new reconstruction algorithm \cite{teo} has been proposed to improve the resolution whatever the angle $\uptheta$. This work describes the implementation of the algorithm to the $\upmu$-RWELL \cite{jinst15}.
\section{The charge centroid (CC) method}
\begin{figure}[!ht]
    \centering
    \includegraphics[width=0.60\textwidth]{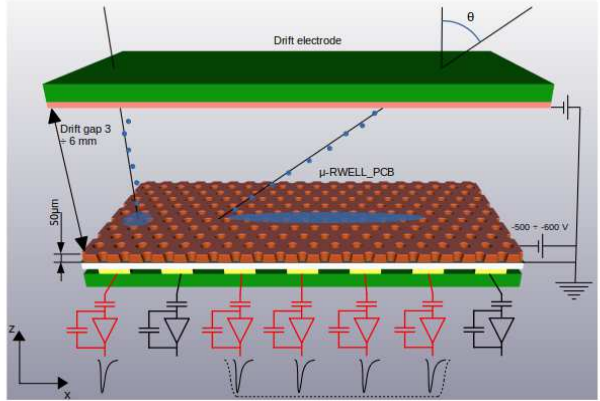}
    \caption{A simplified sketch showing how the non orthogonal tracks affect the number of fired strips.}
    \label{sketch}
\end{figure}

For a detector equipped with a strip-segmented readout and instrumented with analog FEE, when a set of strips is fired the position of the track can be computed as
\begin{equation}
X_{CC}=\frac{\sum x_{k}q_{k}}{\sum q_{k}}    
\end{equation}
where $x_{k}$ is the coordinate of the $k$-th strip and $q_{k}$ is its integrated charge. The uncertainty associated to this position is strongly dependent on the impinging angle ($\uptheta$) of the track (fig.~\ref{sketch}). To overcome this issue a new algorithm has been recently proposed. 

\section{The $\upmu${u}TPC algorithm}
\begin{figure}
	\centering
	\begin{minipage}[t]{.48\textwidth}
		\centering
		\vspace{0pt}
		\includegraphics[scale=0.4]{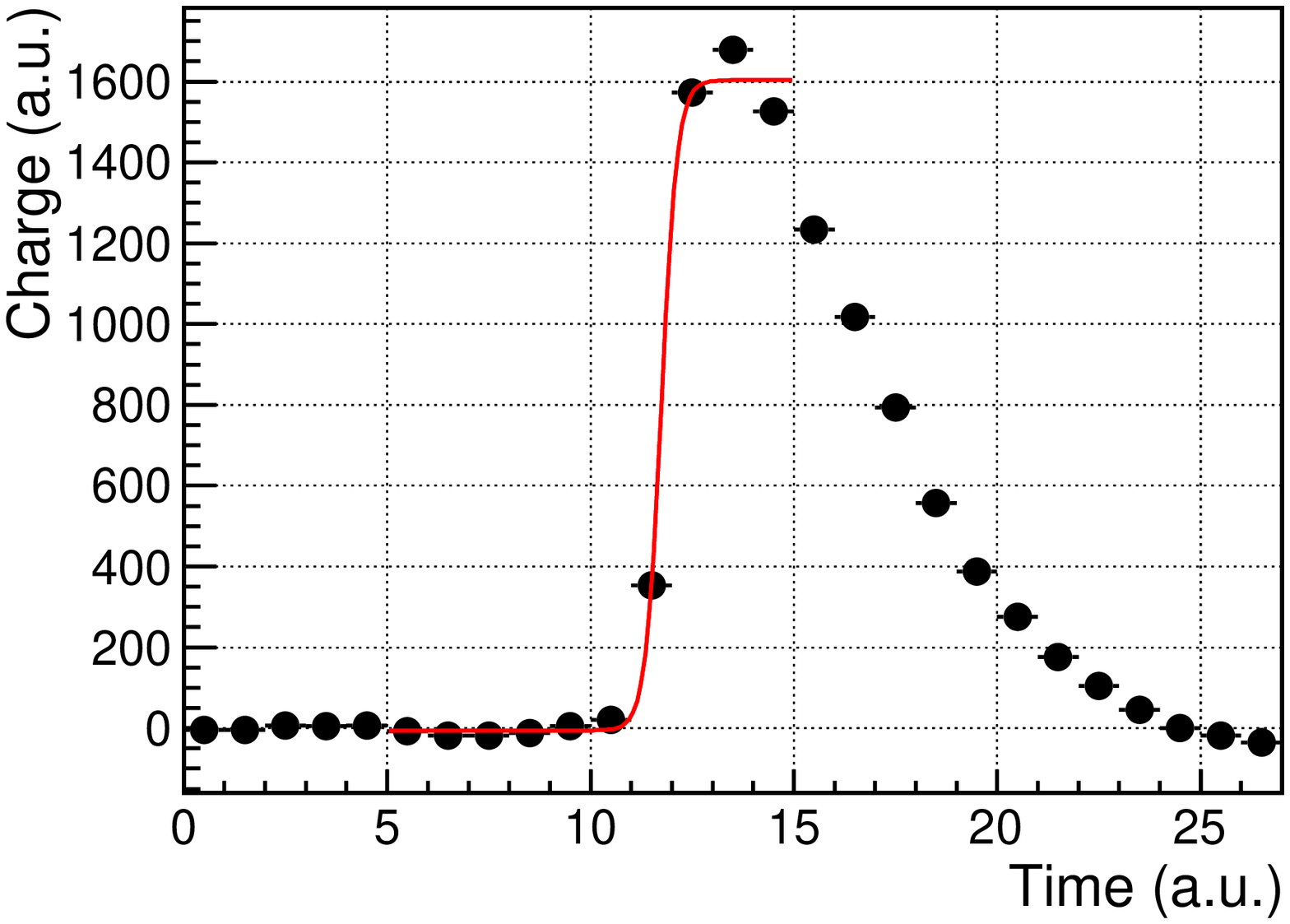}
        \caption{Integrated charge as a function of the sampling time, with the fitting Fermi-Dirac function.}
        \label{qvstime}
	\end{minipage}\hfill
	\begin{minipage}[t]{.48\textwidth}
		\centering
		\vspace{0pt}
		\includegraphics[scale=0.4]{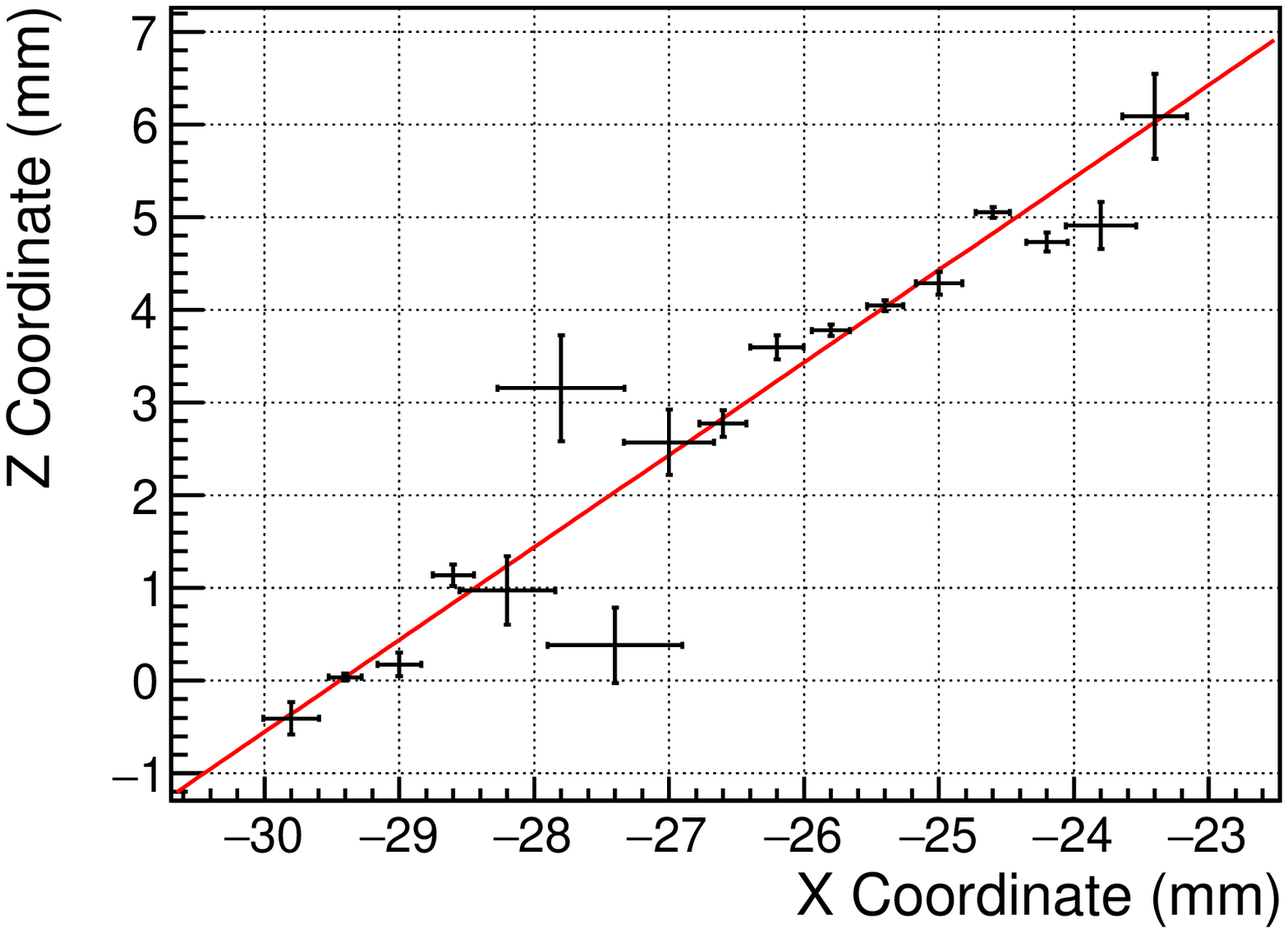}
        \caption{Example of a $45^{\circ}$ track segment reconstruction using the $\upmu$TPC algorithm. The line is the linear fit.}
        \label{example}
	\end{minipage}
\end{figure}

The idea developed for the ATLAS MicroMegas of the New Small Wheels \cite{teo, iaco}, and also implemented on the BESIII cylindrical GEM \cite{ferrara-utpc,ferrara-cgem}, is to reconstruct a track segment inside the detector conversion gap rather than a single hit. The procedure is inspired to the Time Projection Chamber (TPC) concept \cite{nygren1974, nygren1978} 
exploiting the analog readout of the signals.
The electrons created by the ionizing particle drift towards the amplification region. 
By the measurement of electrons arrival time
and knowing their drift velocity in the gas mixture, the position of the
ionization clusters can be localized in the chamber. A fit to these clusters provide the 3D trajectory of the ionizing particle. 
In our case the readout is segmented in 1D strips, so only a reconstruction in what we define the $x-z$ plane (fig.~\ref{sketch}) is available.
The fired strips represent the projection of the track on the readout and each center is the $x$ coordinate of the corresponding ionization.
These hits are recorded at different times $t_{k}$, depending on the distance of the ionization electrons from the readout plane. 
Applying the simple formula 
\begin{equation}
z_{k}=v_{drift}\cdot\left(t_{k}-t_{0}\right)
\label{z}
\end{equation} 
the $z$ position of the $k$-th cluster can be computed. 
The formula \ref{z} exploits the good uniformity of the drift field in MPDG detectors, so that the velocity of the electrons can be considered constant. The drift velocity $v_{drift}$ of the electrons as a function of the drift field in several gas mixtures can be found in literature. Anyway a fast tool to catch these measurements is the MAGBOLTZ \cite{magb} routine called by the GARFIELD gas detector simulation program \cite{garf}.
The $t_{0}$ is the common trigger time. It is crucial to define the best value for $t_{k}$. In our case, using the FEE APV25 \cite{apv}, the integrated charge is sampled every 25 ns (fig.~\ref{qvstime}). The leading edge of this plot is fitted with a Fermi-Dirac function and its flex point is taken as the $t_{k}$ for the eq. \ref{z}. 
In fig.~\ref{example} it is shown the track segment reconstruction of an event using this algorithm. The error bars on the x axis basically account for the strip pitch and for the fraction of the total charge collected on the strip (errors are increased for small charges possibly associated to charge induction); the error bars on the z axis are propagated from the time measurement uncertainty. Another possible choice for the reconstructed point errors is stated in \cite{iaco}.

The x coordinate of the event is interpolated from the linear fit, taking the coordinate of the track at the middle plane of the drift space, following the approach of \cite{iaco} and \cite{ferrara-utpc}.

\section{Results}
\begin{figure}
	\centering
	\begin{minipage}[t]{.48\textwidth}
		\centering
		\vspace{0pt}
		\includegraphics[scale=0.26]{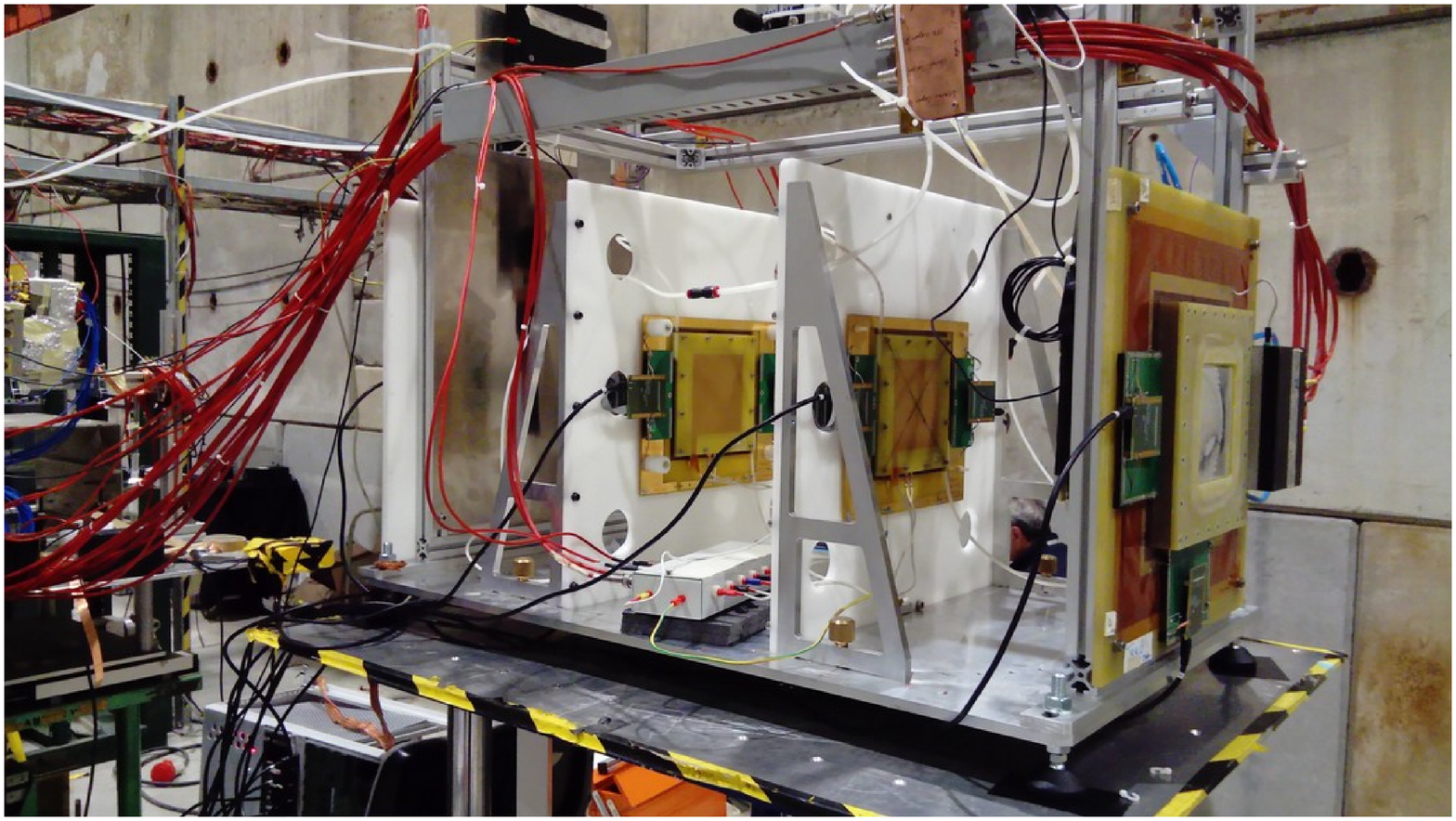}
		\caption{Experimental setup.}
		\label{setup}
	\end{minipage}\hfill
	\begin{minipage}[t]{.48\textwidth}
		\centering
		\vspace{0pt}
		\includegraphics[scale=0.15]{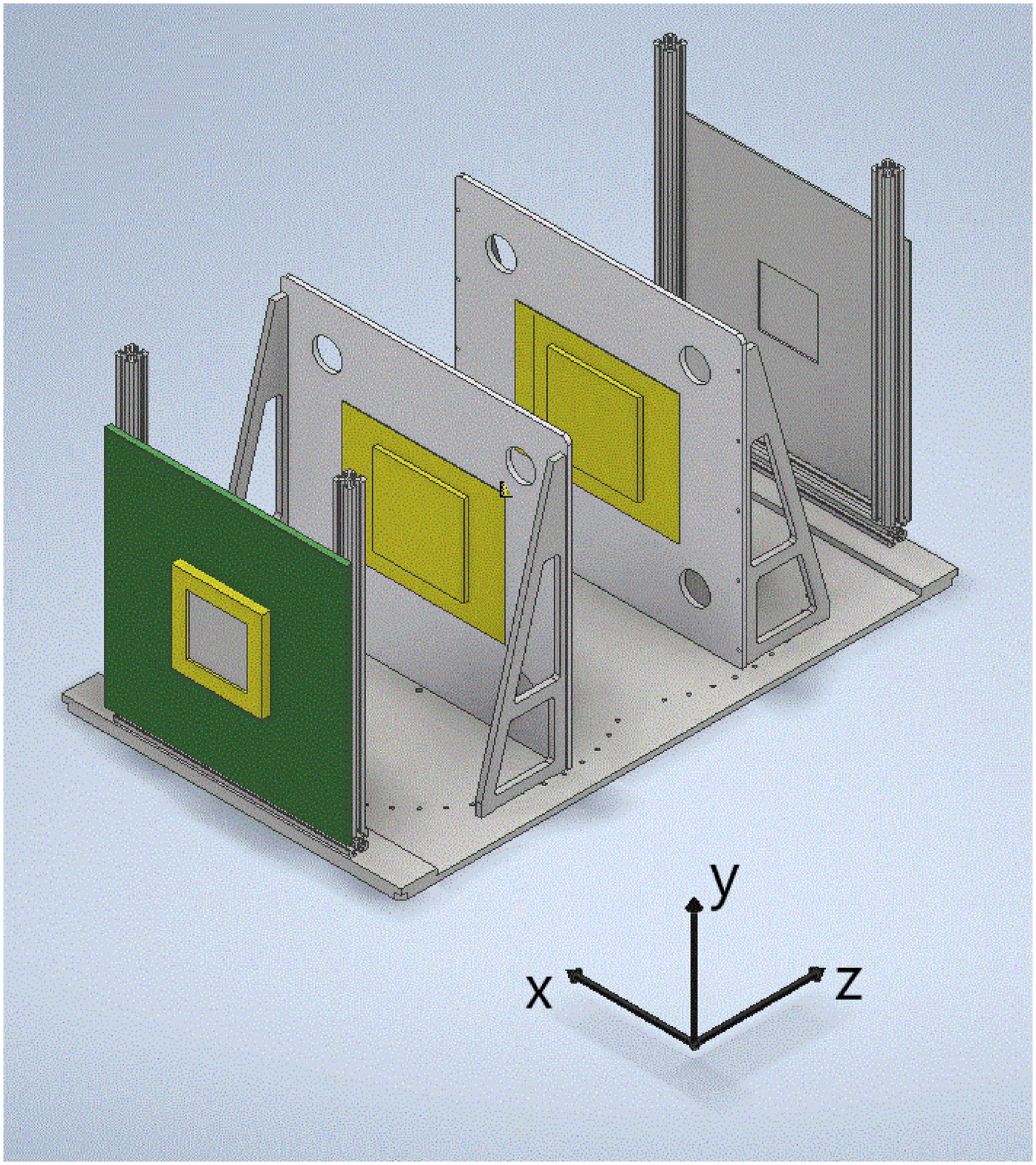}
		\caption{Sketch of the setup with the coordinate system.}
	\end{minipage}
\end{figure}

\begin{figure}
    \centering
    \includegraphics[scale=0.4]{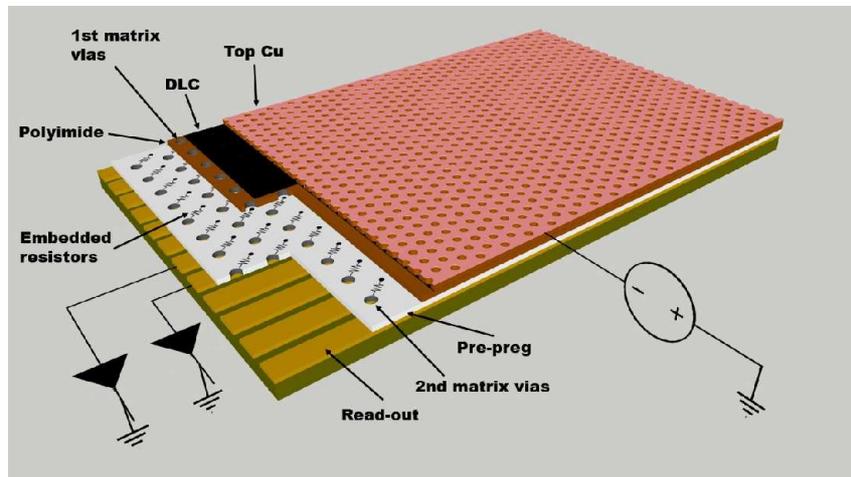}
    \caption{Sketch of the Double Resistive Layer $\upmu$-RWELL with embedded resistors.}
    \label{DRL}
\end{figure}

Measurements of the space resolution of the $\upmu$-RWELL where only the charge centroid method has been applied are reported in \cite{nim}. According to those results for the following tests DLC foils with resistivity ranging between 60 and 200~M$\Omega/\Box$ have been selected for the realization of the detectors. The $\upmu$TPC algorithm has been used with $\upmu$-RWELLs during a test beam at H8-SPS CERN with a 150~GeV/c muon beam. Two GEM detectors (fig.~\ref{setup}) have been used to select fully reconstructed tracks in order to clean. Two $\upmu$-RWELLs have been installed on rotating plates so that the beam could form different angles with
respect to the normal to the electrodes. 
The $\upmu$-RWELLs used in the test (fig.~\ref{DRL}) are derivation of the DRL layout \cite{drl}: two metallic vias matrices connect two resistive stages to the readout plane for the grounding. The vias density is typically $\leq$~1~cm$^{-2}$. The first stage is a DLC layer, while the second is made of $\sim 5$ mm long resistors screen-printed on a substrate. 
The detectors are equipped with a strip-segmented readout (400 $\upmu$m pitch), operated at a gain of 5000 with readout APV25 front-end electronics and flushed with Ar:CO$_{2}$:CF$_{4}$ 45:15:40 gas mixture.

The space resolution can be extracted from the distribution width of the residuals ($\sigma_{res}$), that are defined as the difference between the coordinates reconstructed by the two $\upmu$-RWELLs. Indeed assuming the same contribution, the $\upmu$-RWELL space resolution is obtained as $\sigma_{x} = \sigma_{res}/\sqrt{2}$. For sake of simplicity in this paper all the plots showing the residual distribution are scaled by a factor of $1/\sqrt{2}$ in order to directly give the detector space resolution. 
The residuals are evaluated and studied for both the charge centroid and for the $\upmu$TPC reconstruction.

In order to take into account the presence of tails, we fit the data with the sum of two gaussian curves, eq.~\ref{eq:doublegaussfunc}. The width of the residuals is defined as its standard deviation, eq.~\ref{eq:doublegausssigma}.
This is a slightly different approach with respect to the analysis reported for MicroMegas (\cite{iaco}). A discussion about the two methods is shown in appendix~\ref{app:comparison}.

\begin{eqnarray}
    f(x) & = & A e^{-\frac{1}{2}\left(\frac{x-\mu_1}{\sigma_1}\right)^2}+B e^{-\frac{1}{2}\left(\frac{x-\mu_2}{\sigma_2}\right)^2}\label{eq:doublegaussfunc}\\
    \sigma & = & \frac{1}{(A\sigma_1+B\sigma_2)}\sqrt{A^2 \sigma_1^4+B^2\sigma_2^4+A B \sigma_1 \sigma_2 \left((\mu_1-\mu_2)^2+\sigma_1^2+\sigma_2^2\right)}\label{eq:doublegausssigma}
\end{eqnarray}

It has been necessary to evaluate and to reduce the systematic effects present in the measurements, among which the most important are the dependency on the $x$ coordinate and the beam divergence. This must be done for both CC and $\upmu$TPC algorithm. In the following the correction of the residuals, reconstructed with the $\upmu$TPC algorithm, as a function of the $x$ coordinate is shown as an example of this procedure.
The detectors have been operated with a drift field of 1~kV/cm and an impinging angle ($\uptheta$) of $30^\circ$.

\begin{figure}
	\centering
	\begin{minipage}[t]{.48\textwidth}
		\centering
		\vspace{0pt}
		\includegraphics[scale=0.4]{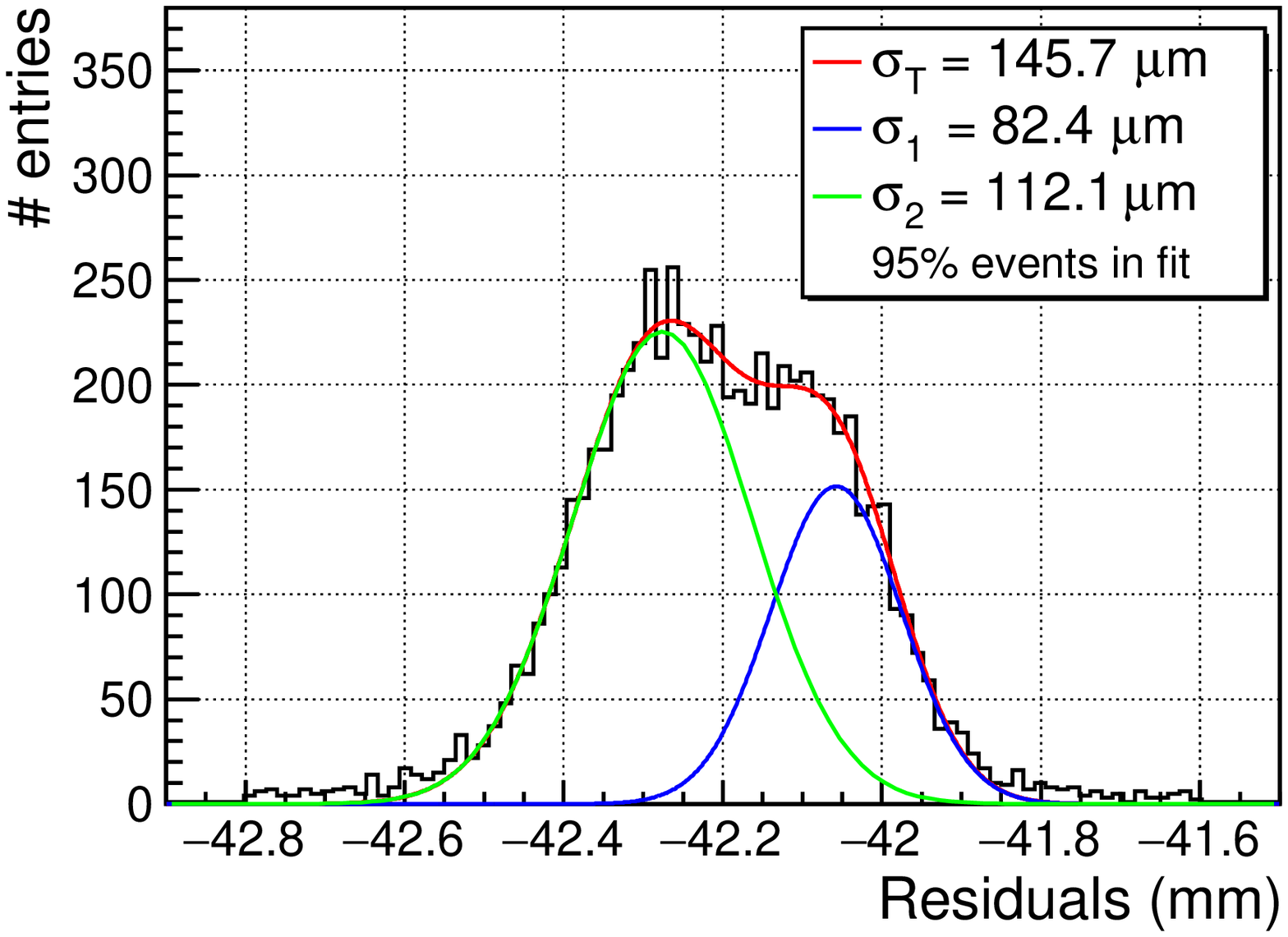}
		\caption{Residuals distribution before any correction.\label{res_0}}
	\end{minipage}\hfill
	\begin{minipage}[t]{.48\textwidth}
		\centering
		\vspace{0pt}
		\includegraphics[scale=0.4]{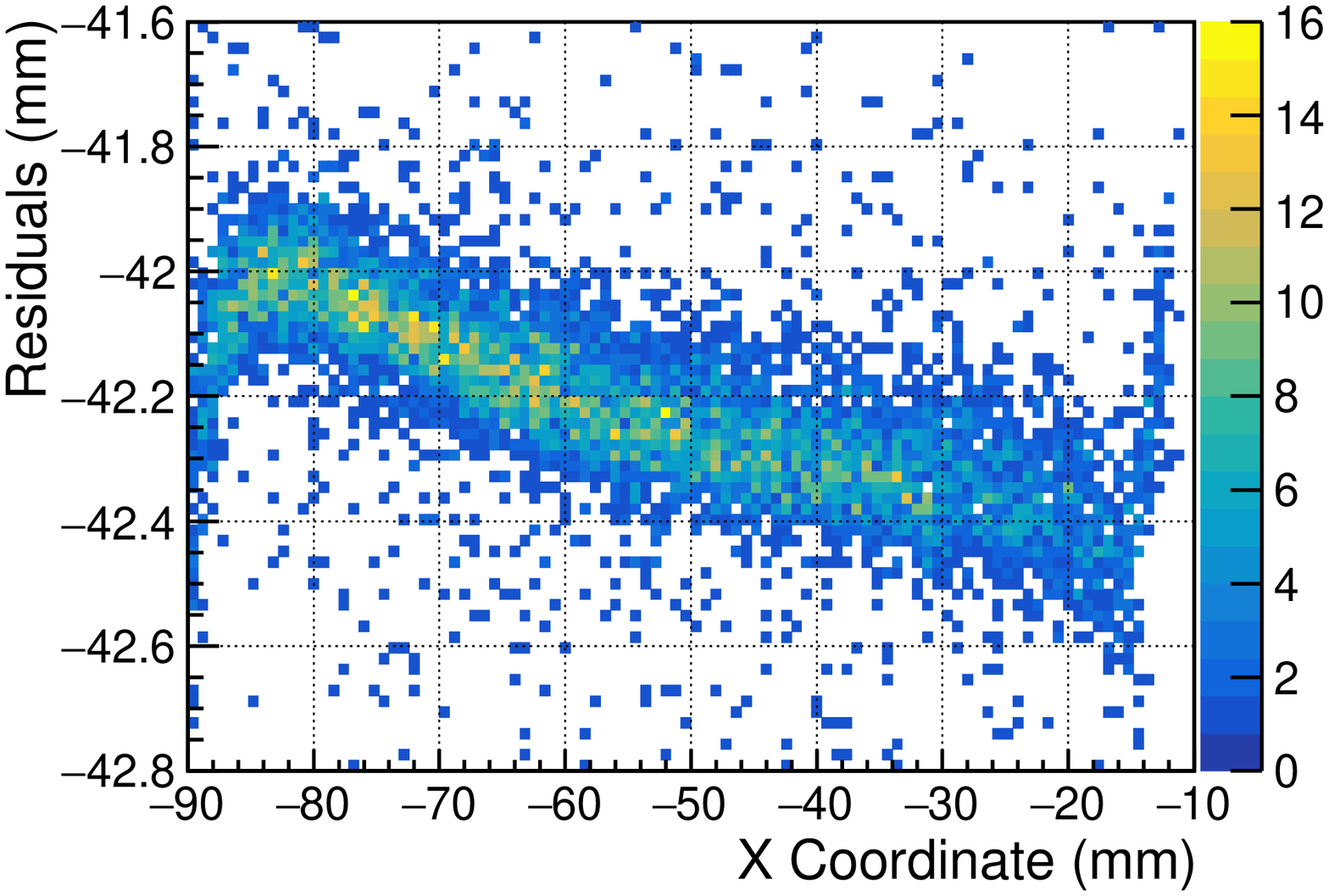}
		\caption{Residuals as a function of the $x$ coordinate.\label{res_vs_x}}
	\end{minipage}
\end{figure}

\begin{figure}
     \subfloat[Profile before the correction.\label{prof_res_vs_x}]
     {
       \includegraphics[width=0.48\textwidth]{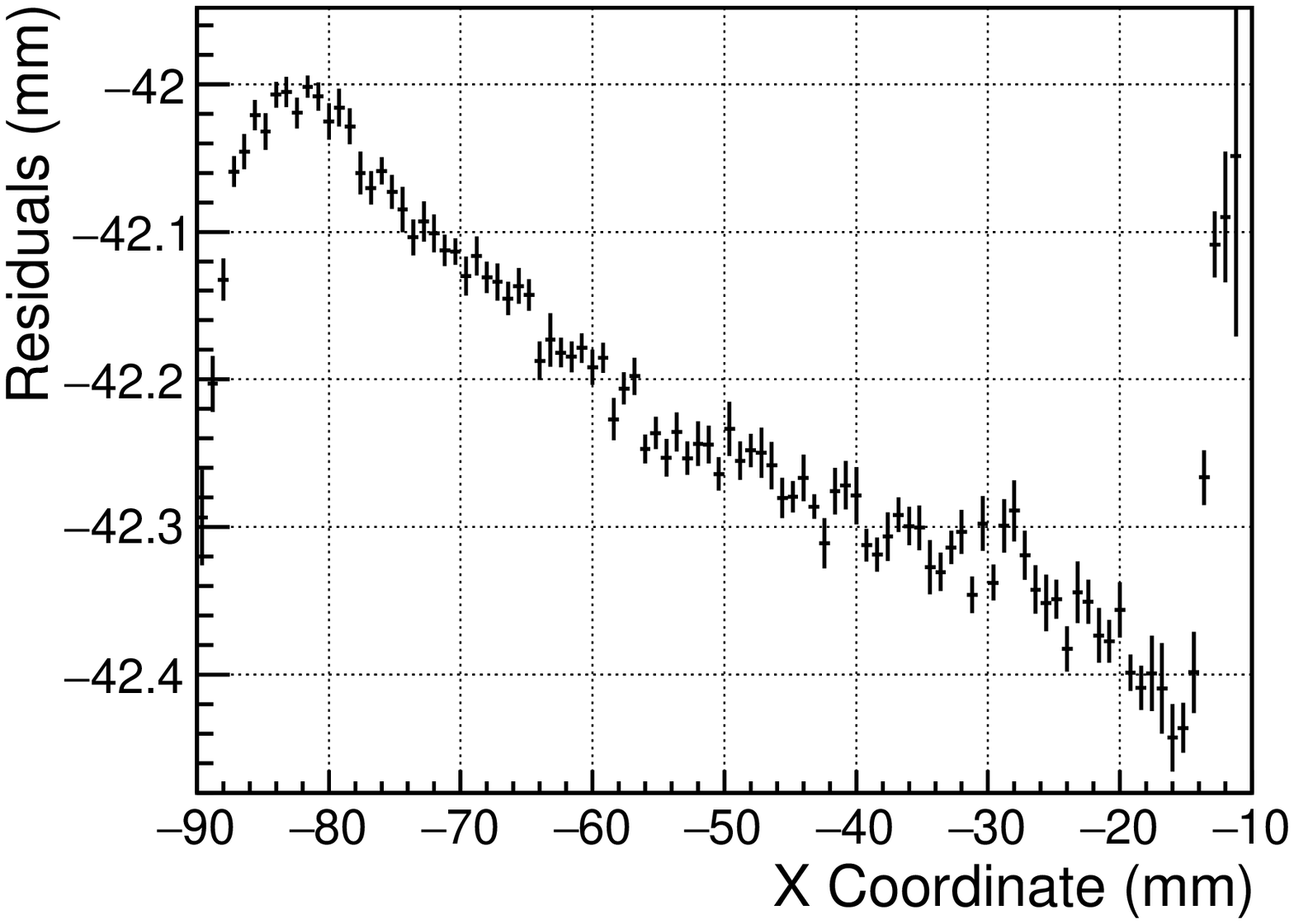}
     }
     \hfill
     \subfloat[Profile after the correction.\label{prof_res_cor}]
     {
       \includegraphics[width=0.48\textwidth]{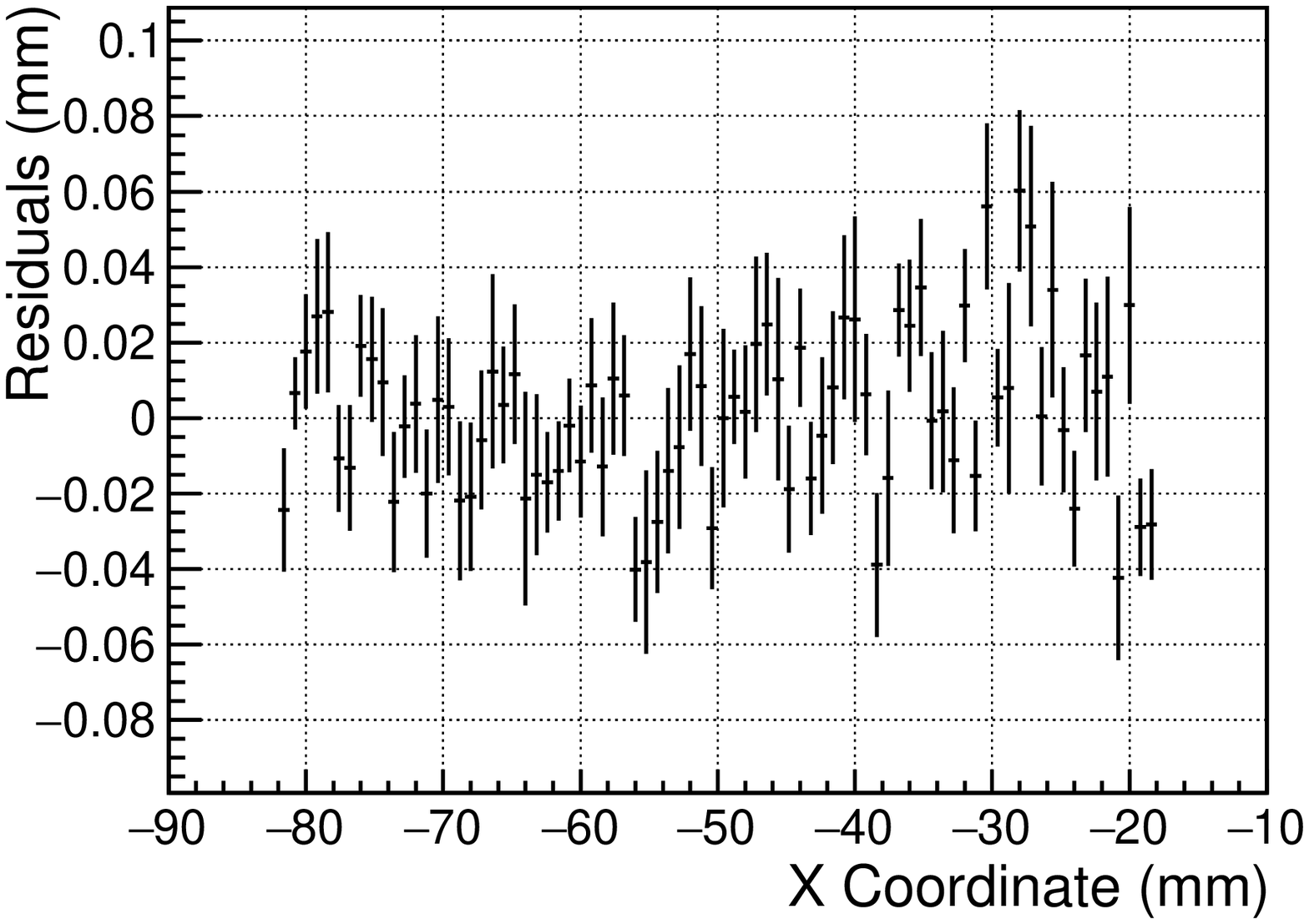}
     }
     \caption{Dependencies of the residuals distribution on the $x$ coordinate.}
\end{figure}

\begin{figure}
	\centering
	\begin{minipage}[t]{.48\textwidth}
		\centering
		\vspace{0pt}
		\includegraphics[scale=.4]{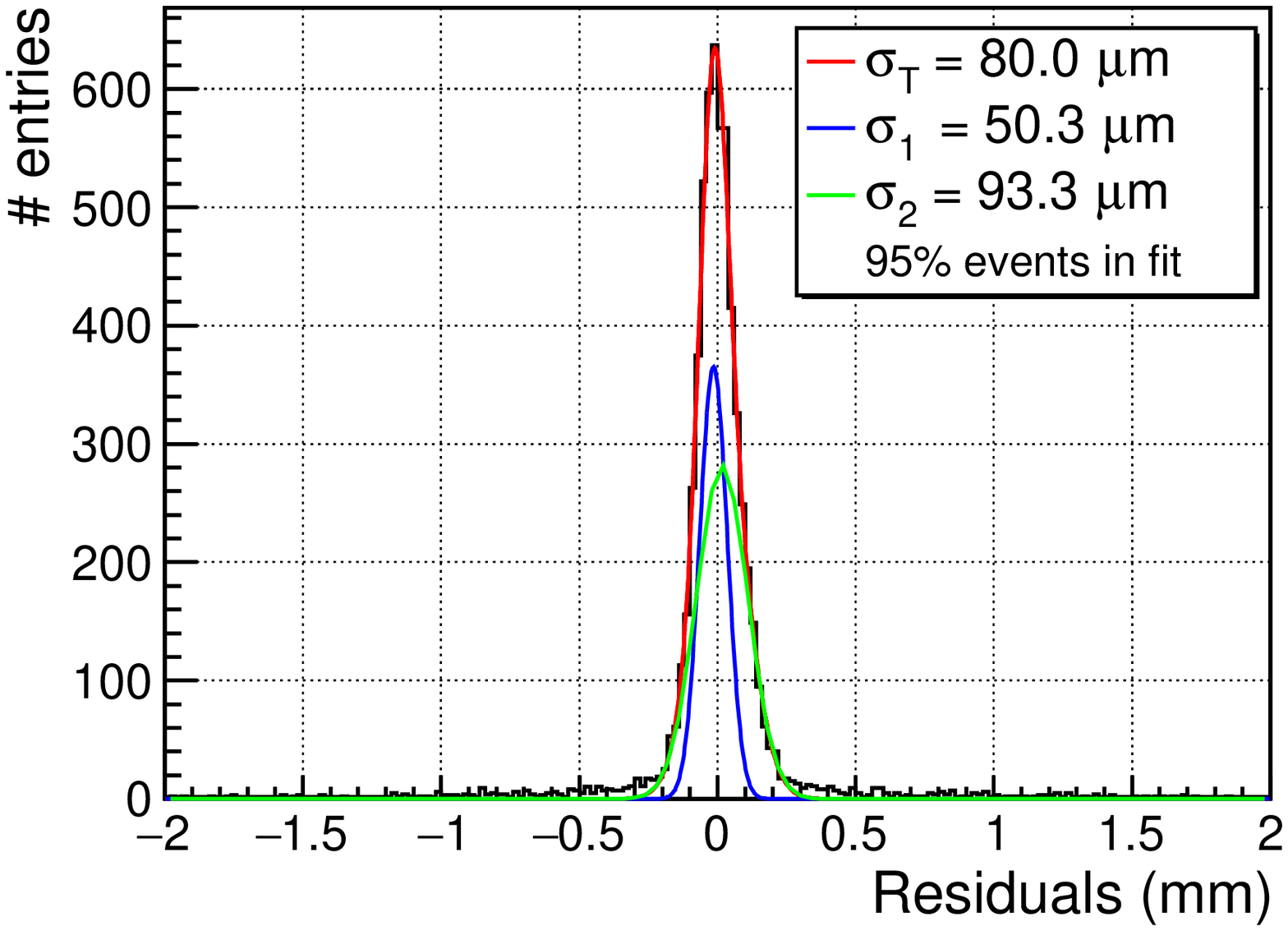}
		\caption{Residuals distribution after the $x$ coordinate correction.\label{res_1}}
	\end{minipage}\hfill
	\begin{minipage}[t]{.48\textwidth}
		\centering
		\vspace{0pt}
		\includegraphics[scale=.4]{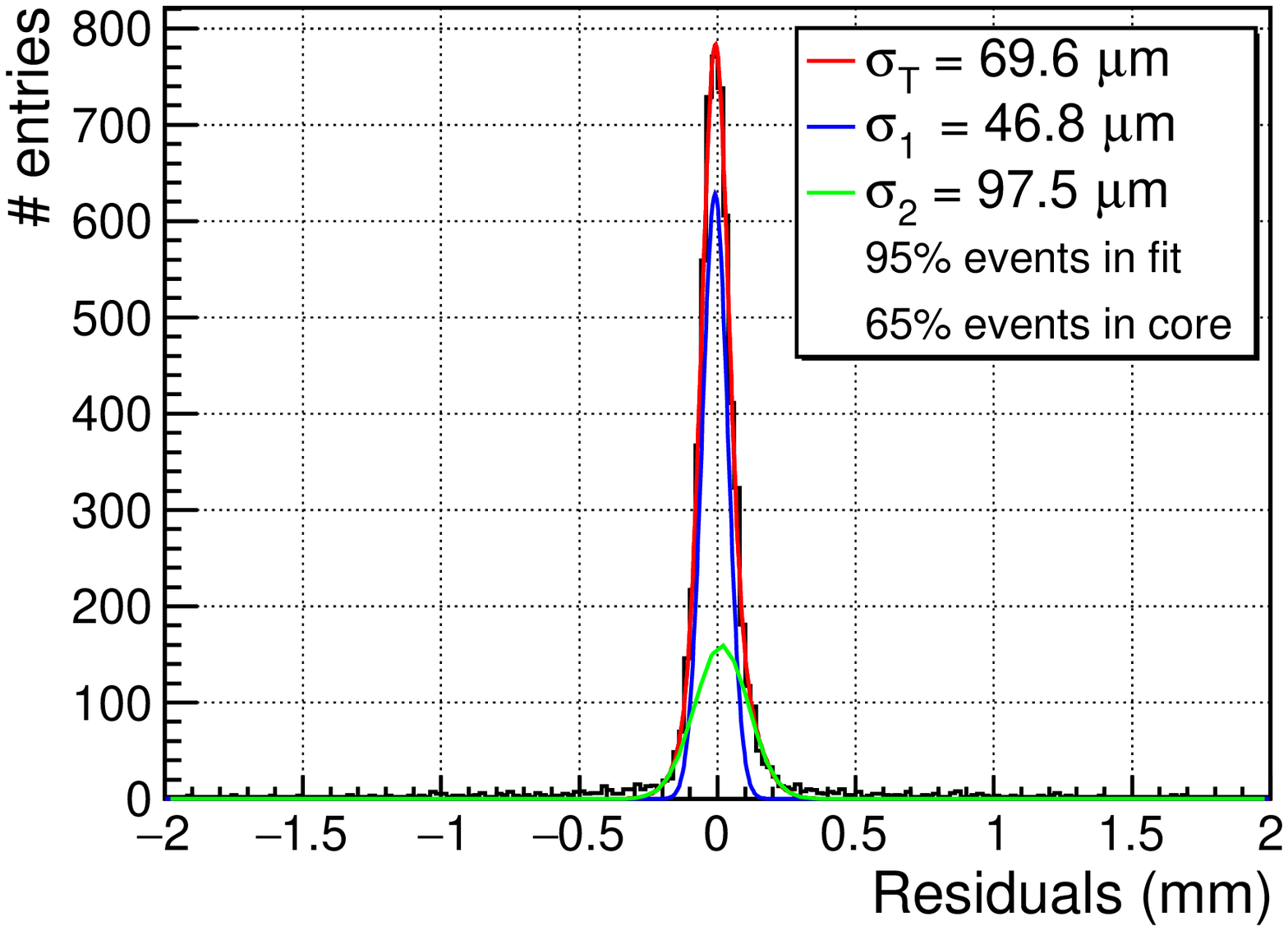}
		\caption{Residuals distribution after all the systematic corrections.\label{residuals}}
	\end{minipage}
\end{figure}

In fig.~\ref{res_0} is shown the raw residual distribution. Plotting it as a function of the $\upmu$TPC-reconstructed $x$ coordinate, fig.~\ref{res_vs_x}, it is visible a clear dependence, evaluated through a profile, fig.~\ref{prof_res_vs_x}. Such profile is then fitted with a suitable polynomial and the residuals are corrected accordingly (figs.~\ref{prof_res_cor},\ref{res_1}).
The residual distribution after all the corrections is shown in fig.~\ref{residuals}.
The distributions are fitted with function~\ref{eq:doublegaussfunc} over 95\% of the events in the histogram.

\begin{figure}[!ht]
     \subfloat[CC spatial resolution.\label{conf_cc}]
     {
       \includegraphics[width=0.48\textwidth]{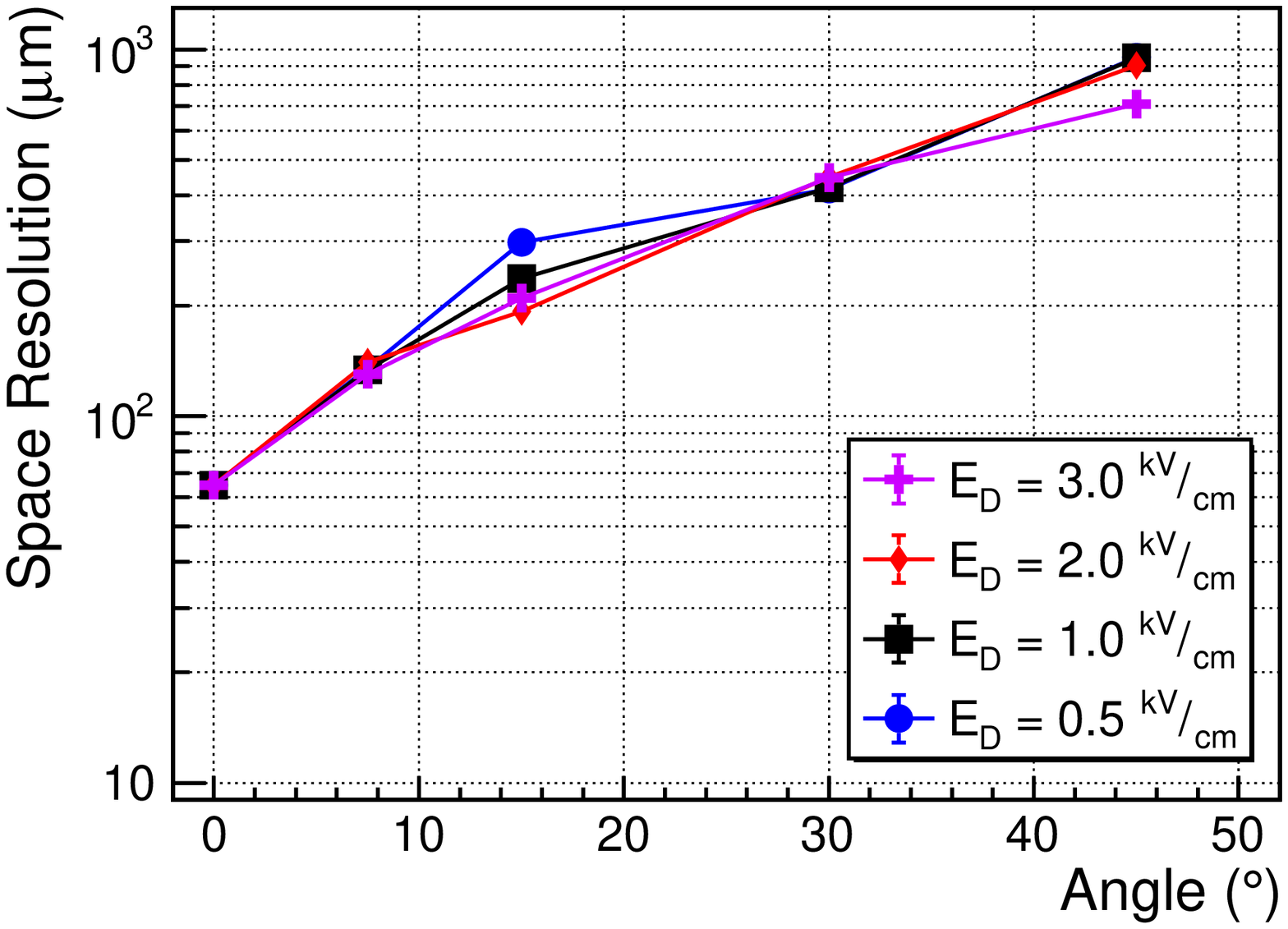}
     }
     \hfill
     \subfloat[$\upmu$TPC spatial resolution.\label{conf_tpc}]
     {
       \includegraphics[width=0.48\textwidth]{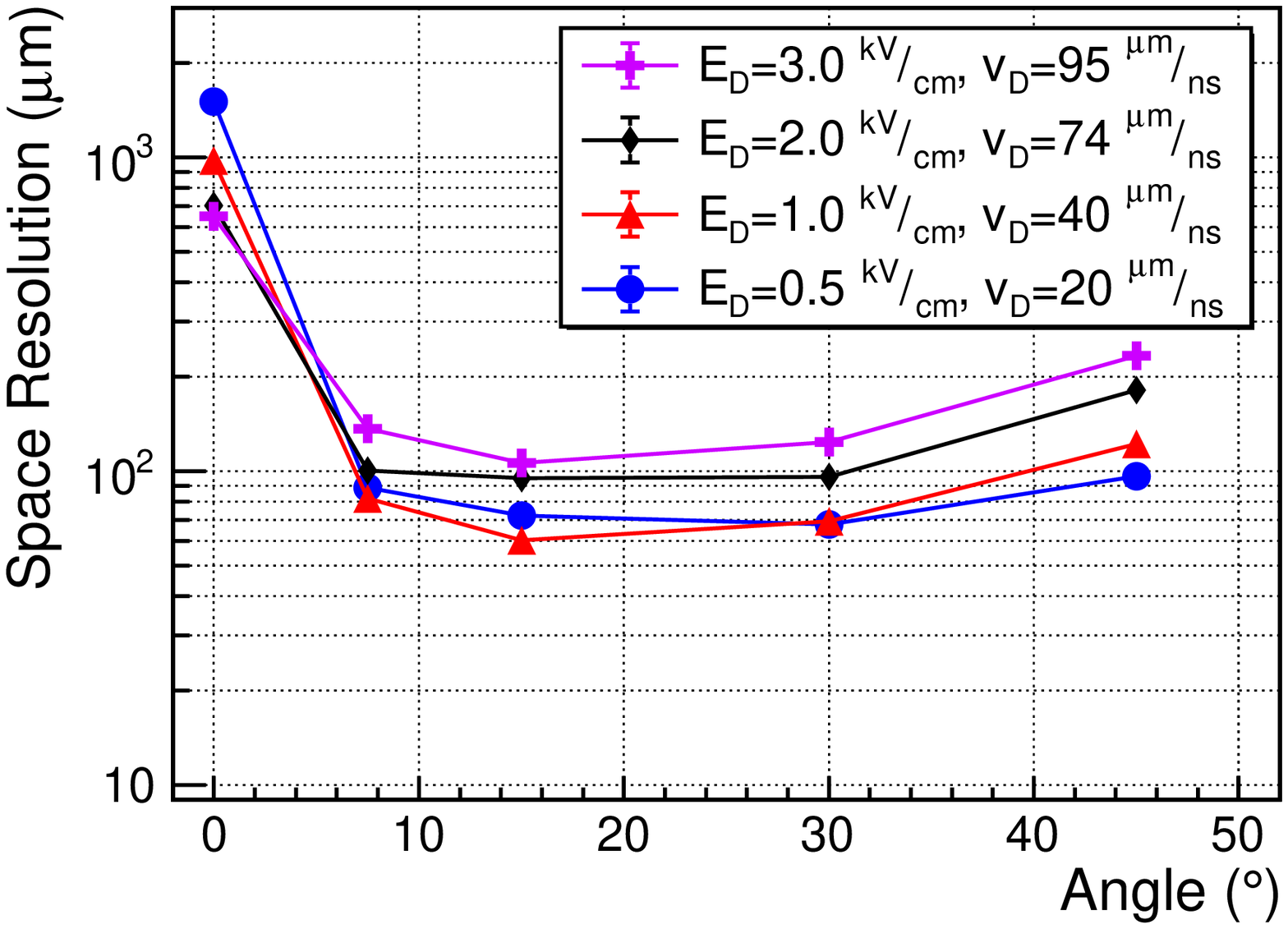}
     }
     \caption{The results of the two reconstruction algorithm, over a large angle, for various drift field values (E$_{D}$).}
\end{figure}

The space resolution has been evaluated at different $\uptheta$ using both CC and $\upmu$TPC methods. As expected for orthogonal tracks the CC provides better results while increasing the angle they quickly worsen, fig.~\ref{conf_cc}. Viceversa the $\upmu$TPC algorithm shows a better behavior for large angles than for small ones (fig.~\ref{conf_tpc}) for wich the longer projected track segment on the readout plane corresponds to a larger number of points to be fitted.

Since the $\upmu$TPC method depends on the drift velocity of the ionization electrons in the gas mixture, and consequently on the drift field, a study at different drift fields has been performed (fig.~\ref{conf_tpc}).
For our gas mixture the electron drift velocity increases with the drift fields, in the range 0.5$\div$3~kV/cm, \cite{miscela}.
A smaller drift velocity allows the reconstruction of the $z$ coordinate with a smaller uncertainty, improving the $\upmu$TPC fit.

It is worth noticing that in an experiment it is not possible to determine which algorithm is the best since the track inclination is known just \textit{a posteriori}.
Just to estimate the effect of this combination on the global space resolution we consider the following trivial relation:
\begin{equation}
\frac{1}{\sigma^{2}_{comb}}=\frac{1}{\sigma^{2}_{CC}}+\frac{1}{\sigma^{2}_{\mu TPC}}\label{eq:combined}
\end{equation}

\begin{figure}[!ht]
     \subfloat[Comparison of the two reconstruction algorithms at a drift field E$_{D}$=1~kV/cm.\label{1utpc}]
     {
       \includegraphics[width=0.48\textwidth]{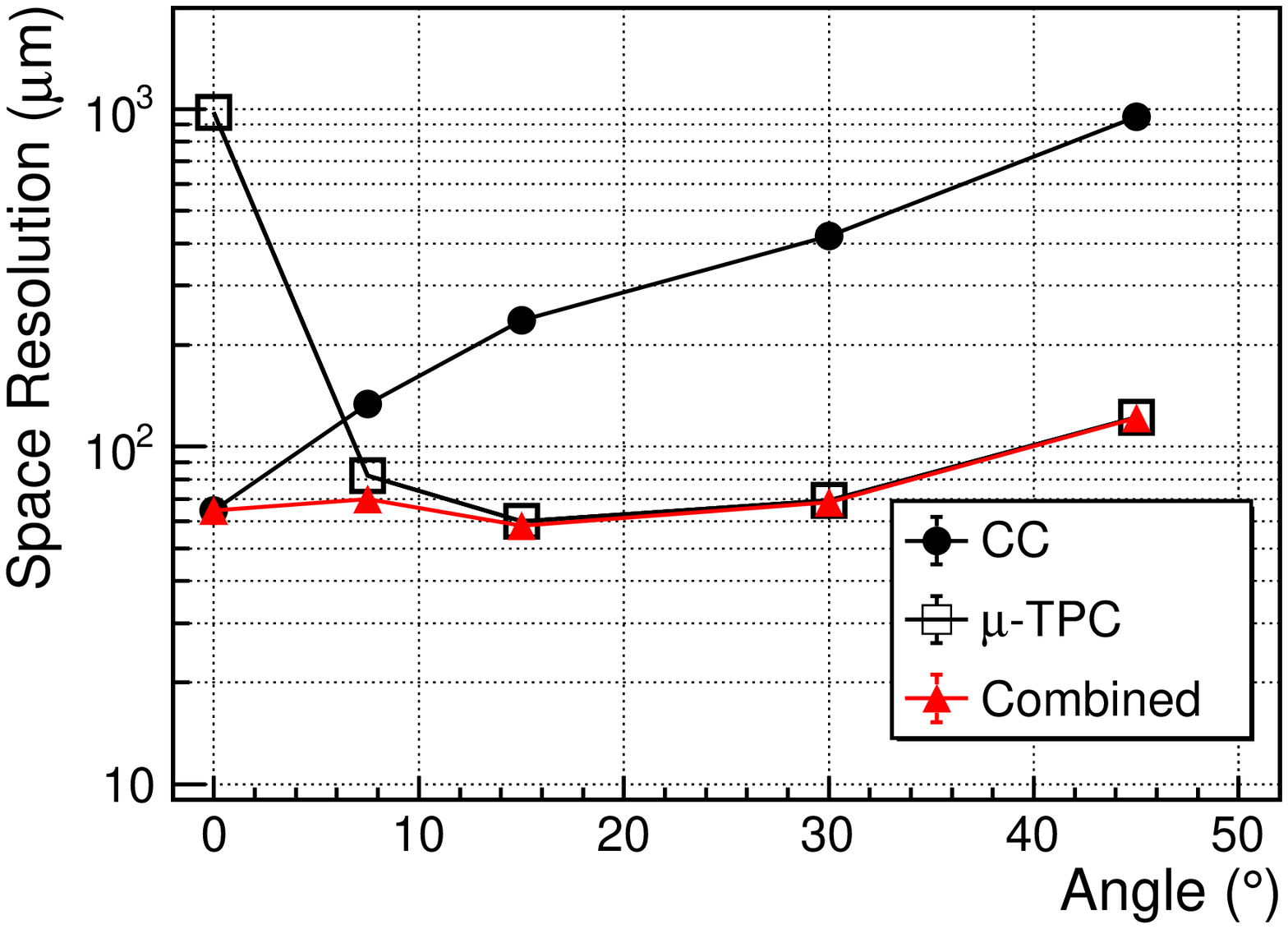}
     }
     \hfill
     \subfloat[Combined space resolution at different drift fields (E$_{D}$) with corresponding drift velocity.\label{drift_comp}]
     {
       \includegraphics[width=0.48\textwidth]{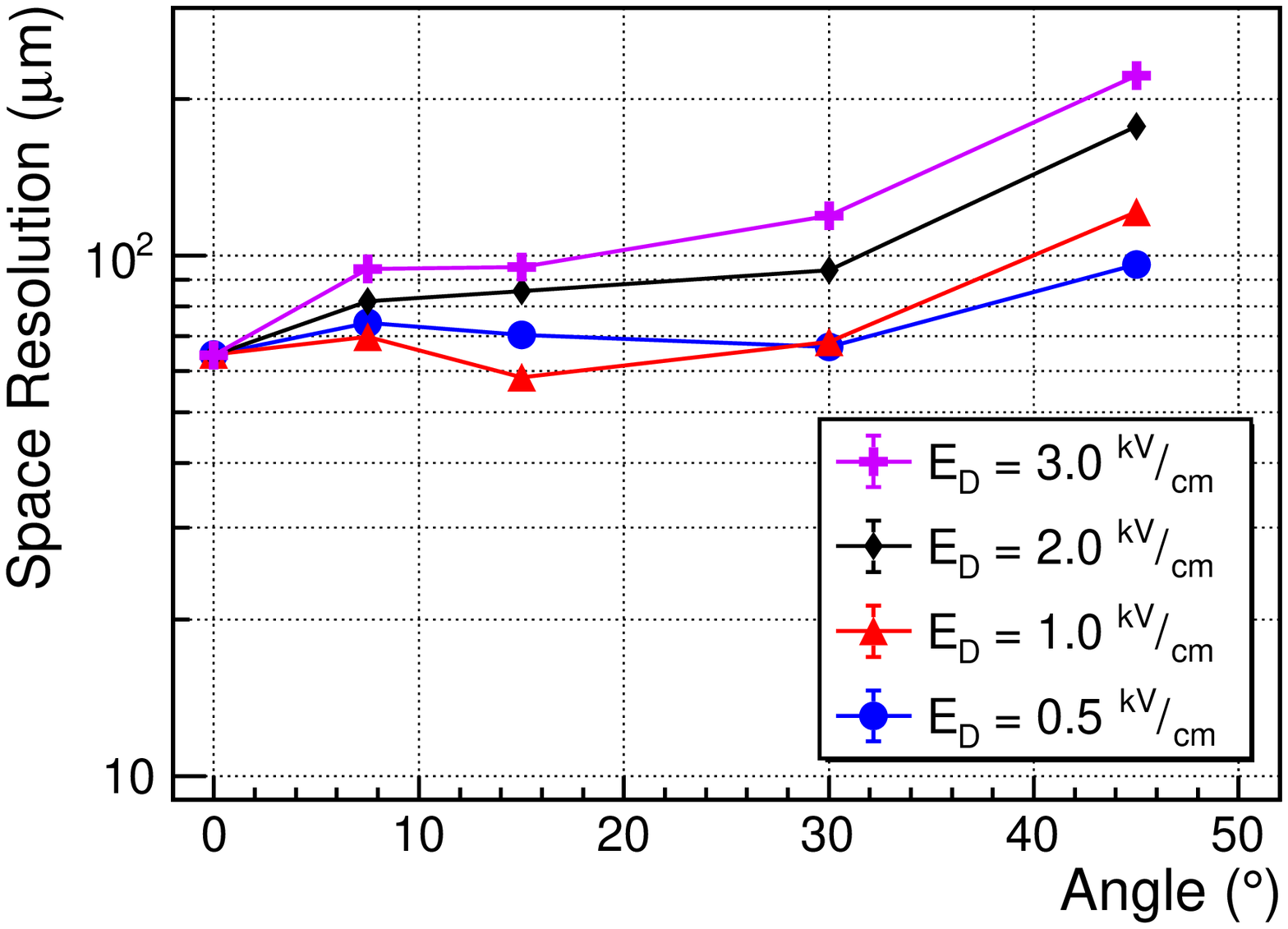}
     }
     \caption{Results from the CC and $\upmu$TPC methods.}
\end{figure}

In fig.~\ref{1utpc} the resolutions for both CC and the $\upmu$TPC are compared and displayed along the combined resolution from eq.~\ref{eq:combined}.

To complete our study we report in fig.~\ref{drift_comp} the combined space resolution at different drift fields.
The combination of the two algorithms results in space resolutions below 100~microns over a large set of angles $\uptheta$., for fields up to 2~kV/cm.

For orthogonal tracks the CC resolution prevails in the combination and it does not depend on the drift field in this range. 

\section{Conclusions}
The $\upmu$TPC method has been succesfully implemented for the tracks reconstruction on the $\upmu$-RWELL. By combining the $\upmu$TPC algorithm with the charge centroid, an almost uniform space resolution over a wide range of track incidence angles is reached. At low drift field the measured space resolution is improved reaching values below 80~$\upmu$m, reaching a minimum of 60$\upmu$m.

\appendix
\section{Consideration upon the double gaussian fit}\label{app:comparison}
As previously stated, equations~\ref{eq:doublegaussfunc} and \ref{eq:doublegausssigma} were used to estimate the spatial resolution of the $\upmu$-RWELL detectors. There is not an univocal approach to this task, for example in \cite{iaco} the width of the residual distribution, fitted with the same function~\ref{eq:doublegaussfunc}, was defined as
\begin{equation} \label{eq:doublegaussianMM}
    \sigma^2=\frac{V_1\sigma_1^2+V_2\sigma_2^2}{V_1+V_2},
\end{equation}
in which $V_{1,2}$ are the integrals of the two gaussian functions: $V_1=\sqrt{2\pi}A\sigma_1$ and $V_2=\sqrt{2\pi}B\sigma_2$ \cite{iodice}. 
The equation~\ref{eq:doublegausssigma} reduces to \ref{eq:doublegaussianMM} only if the two gaussian curves have the same mean, $\mu_1=\mu_2$, namely for a symmetric residual distribution. The proof follows straightforward:

\begin{eqnarray} \label{eq:doublegaussianMMdemo}
    \sigma^2 &= &\frac{A^2 \sigma_1^4+B^2\sigma_2^4+A B \sigma_1 \sigma_2 \left(\sigma_1^2+\sigma_2^2\right)\pm2AB\sigma_1^2\sigma_2^2}{(A\sigma_1+B\sigma_2)^2}\\
    &=&\frac{(A\sigma_1^2+B\sigma_2^2)^2+AB\sigma_1\sigma_2(\sigma_1-\sigma_2)^2}{(A\sigma_1+B\sigma_2)^2}=\frac{(V_1\sigma_1+V_2\sigma_2)^2+V_1V_2(\sigma_1-\sigma_2)^2}{(V_1+V_2)^2}\\
    &=&\frac{(V_1^2+V_1V_2)\sigma_1^2+(V_2^2+V_1V_2)\sigma_2^2}{(V_1+V_2)^2}=\frac{V_1\cancel{(V_1+V_2)}\sigma_1^2+V_2\cancel{(V_1+V_2)}\sigma_2^2}{(V_1+V_2)^{\cancel{2}}}
\end{eqnarray}

\end{document}